\documentclass[5p,twocolumn,times]{elsarticle}

\usepackage{lipsum}
\usepackage{lineno,hyperref}
\modulolinenumbers[5]
\usepackage[pdftex]{color}
\usepackage[font=footnotesize,labelfont=bf]{caption}
\usepackage[font=footnotesize,labelfont=bf]{subcaption}
\journal{Journal of \LaTeX\ Templates}
\usepackage{colortbl}
\usepackage{xcolor}

\usepackage{amssymb}

\biboptions{compress}

\usepackage[figuresright]{rotating}

\begin{document}

\begin{frontmatter}

\title{Adaptive movement strategy in rock-paper-scissors models}

\address[1]{Escola de Ci\^encias e Tecnologia, Universidade Federal do Rio Grande do Norte\\
Caixa Postal 1524, 59072-970, Natal, RN, Brazil}
\address[2]{Institute for Biodiversity and Ecosystem
Dynamics, University of Amsterdam, Science Park 904, 1098 XH
Amsterdam, The Netherlands}

\author[1]{M. Tenorio}  
\author[1]{E. Rangel} 
\author[1,2]{J. Menezes}

\begin{abstract}
Organisms may respond to local stimuli that benefit or threaten their fitness. The adaptive movement behaviour may allow individuals to adjust their speed to maximise the chances of being in comfort zones, where death risk is minimal. We investigate spatial cyclic models where the rock-paper-scissors game rules describe the nonhierarchical dominance. We assume that organisms of one out of the species can control the mobility rate in response to the information obtained from scanning the environment.
Running a series of stochastic simulations, we quantify the effects of the movement strategy on the spatial patterns and population dynamics. Our findings show that the ability to change mobility to adapt to environmental clues is not reflected in an advantage in cyclic spatial games. The adaptive movement
provokes a delay in the spatial domains occupied by the species in the spiral waves, making the group more vulnerable to the advance of the dominant species and less efficient in taking territory from the dominated species. Our outcomes also show that the effects of adaptive movement behaviour accentuate whether most individuals have a long-range neighbourhood perception.  Our results may be helpful for biologists and data scientists to  
comprehend the dynamics of ecosystems where adaptive processes are fundamental.
\end{abstract}

\end{frontmatter}

\section{Introduction}
\label{sec:int}
In ecology, understanding how animal foraging behaviour depends on the environmental conditions is essential to define conservation strategies and predict the stability of ecosystems \cite{ecology,foraging,butterfly,BUCHHOLZ2007401}. There is plenty of evidence that many animals adapt their movement following environmental clues \cite{adaptive1,adaptive2,Dispersal,BENHAMOU1989375,coping}. For example, organisms use sensory information to react to attractant or repellent species ( e.g., prey or predator species), diminishing their death risk and promoting species territorial dominance \cite{repellent,Causes,MovementProfitable}. Interacting with the environment makes individuals respond appropriately, adapting either their movement speed (kinesis behaviour) or direction (taxis movement) \cite{Motivation1}. In both cases, instead of an aimless motion, individuals are guided by local cues they instinctively interpret or learn to decipher, i.e., the motion pattern is a result of an innate or conditioned behaviour, respectively\cite{kinesisnew1}. Understanding the organisms' adaptive movement strategy has also helped the development of engineering tools with sophisticated mechanisms of artificial intelligence that allow animats to change strategies to survive in hostile scenarios \cite{animats}.

It is well established that spatial interactions are responsible for the stability of many ecosystems \cite{Nature-bio}. For example, experiments with bacteria \textit{Escherichia coli} proved that the cyclic dominance among strains maintain biodiversity as long as individuals interact locally \cite{Coli,bacteria,Allelopathy}. 
As observed in the experiment with bacteria \textit{Escherichia coli}, the organism of three strains interact according to the rules of the cyclic, nonhierarchical rock-paper-scissors game \cite{Coli,bacteria,Allelopathy}. In this popular game,  
 scissors cut paper, paper wraps rock, rock crushes scissors, describing the selection interaction among species\cite {Directional1, Directional2} - the same cyclic dominance has been reported in systems of lizards and coral reefs \cite{lizards,Extra1}.
This has motivated many authors to propose a number of numerical models based on stochastic simulations of the spatial rock-paper-scissors game where individuals may move motivated by attack or defence strategies \cite{Park3,Park2,Reichenbach-N-448-1046,Szolnoki-JRSI-11-0735, Moura, Anti1,anti2,MENEZES2022101606}.

In general, besides selecting one another following the game rules, organisms reproduce and move randomly or directionally if motivated by behavioural strategies \cite{PhysRevE.97.032415,Moura}. For example, it has been claimed that the coexistence probability increases if individuals move forward regions dominated by the species that provide them protection. This is no longer true whether individuals move into regions with individuals they have an advantage in the selection rules. \cite{Moura}. Although mathematical models indicate that biodiversity is strongly affected by the adaptive movement when individuals' dispersal rate depends on the availability of the resources\cite{Korea,kinesisnew2}, there is a lack of results in the stochastic simulations of non-directional behavioural movement in cyclic models.

%%%%%%%%%%%%%%%%%%%%%%%%%%%%%%%%%%%%%%%%%%%%%%%%%%%%%%%%%%%%%%%%%%%%%%%%
\begin{figure}[t]
	\centering
	\includegraphics[width=45mm]{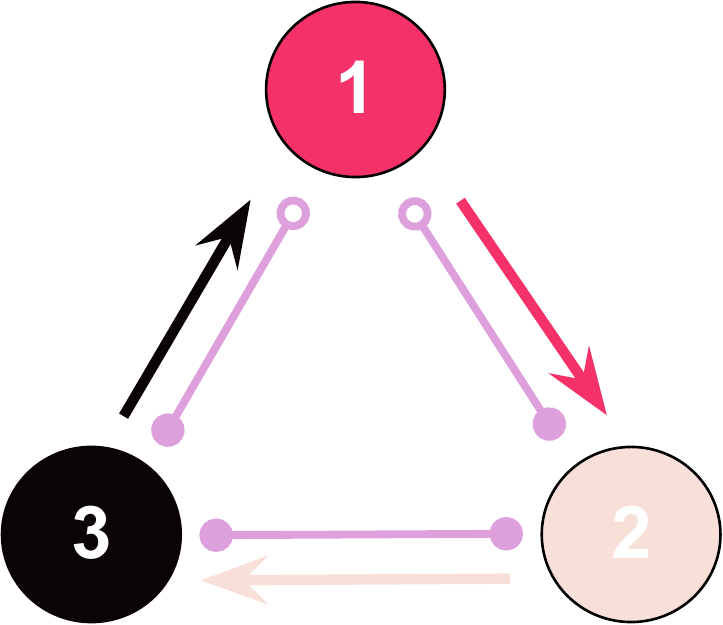}
    \caption{Illustration of selection and mobility interactions. Dark pink, light pink, and black arrows illustrate the cyclic selection interactions, with species $i$ beating species $i+1$. Purple lines indicate that organisms of every species exchange positions during mobility interactions; the open circles indicate that species $1$ can adapt the mobility rate according to local attractiveness.}
  \label{fig1}
\end{figure}
%%%%%%%%%%%%%%%%%%%%%%%%%%%%%%%%%%%%%%%%%%%%%%%%%%%%%%%%%%%%%%%%%%%%%%%
%%%%%%%%%%%%%%%%%%%%%%%%%%%%%%%%%%%%%%%%%%%%%%%%%%%%%%%%%%%%%%%%%%%%%%%%
\begin{figure}[t]
	\centering
	\includegraphics[width=85mm]{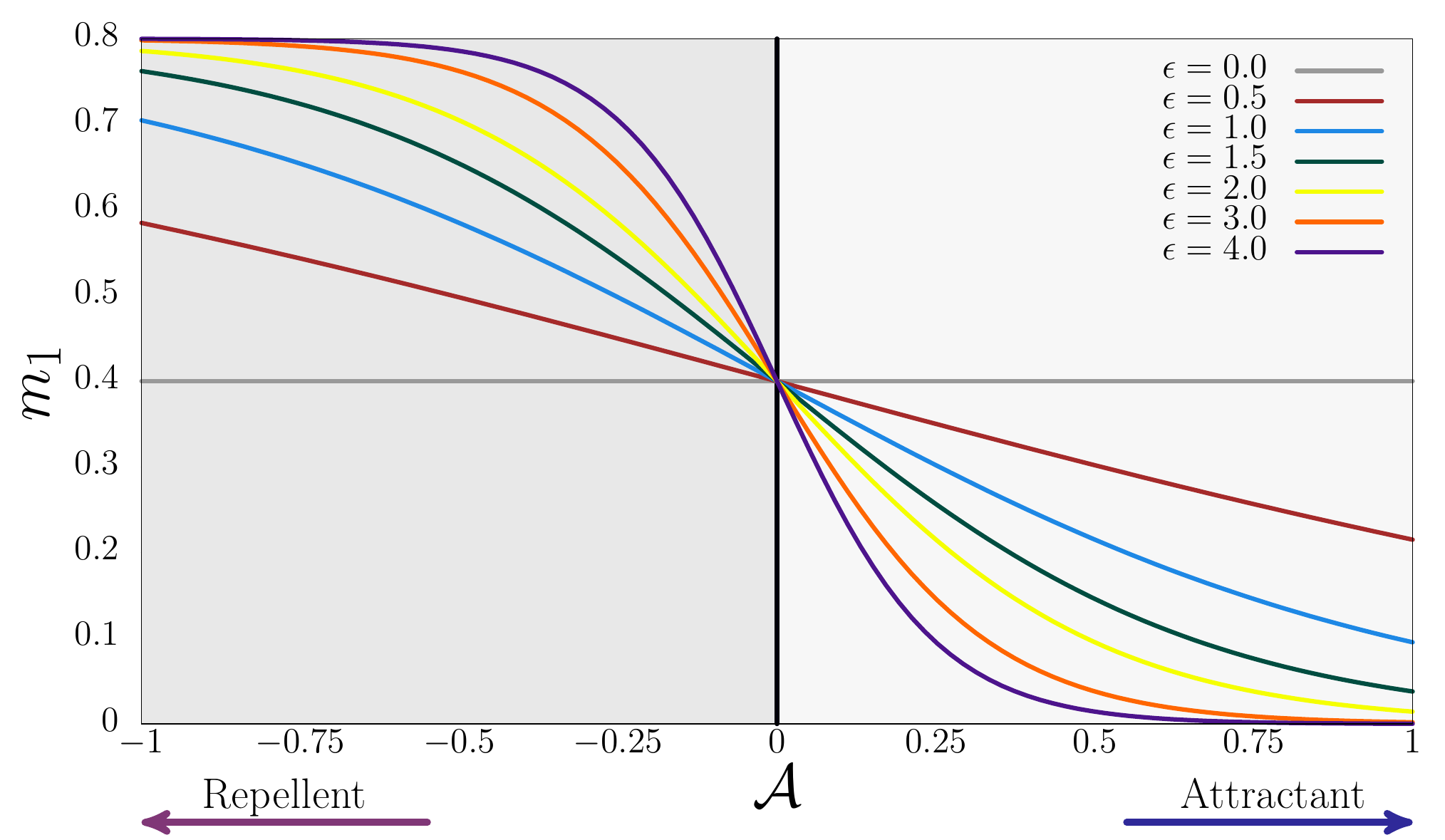}
    \caption{Mobility probability of organisms of species $1$ in terms of the local
attractiveness for $m=0.8$. The grey line indicates that organisms that cannot adapt their movement to local densities ($\epsilon=0.0$) move with constant probability $m=0.4$. For positive values of $\mathcal{A}$ (attractant), the mobility decreases quicker for higher $\epsilon$; for negative $\mathcal{A}$ (repellent), the mobility probability increases more significantly for higher $\epsilon$.}
  \label{fig1b}
\end{figure}
%%%%%%%%%%%%%%%%%%%%%%%%%%%%%%%%%%%%%%%%%%%%%%%%%%%%%%%%%%%%%%%%%%%%%%%

%%%%%%%%%%%%%%%%%%%%%%%%%%%%%%%%%%%%%%%%%%%%%%%%%%%%%%%%%%%%%%%%%%%%%%%

Our goal is to comprehend how adaptive movement strategy interferes with the equilibrium in the dynamics of the spatial organisation in a cyclic system. We perform agent-based stochastic simulations of the rock-paper-scissors game, where individuals of one out of the species adjust their mobility in response to environmental cues. Due to the cyclic dominance of the spatial game, organisms: i) decelerate when in areas dominated by the attractant species - the species they have an advantage in the selection rules; ii) accelerate when in regions occupied mainly by individuals of the repellent species - the species they have an advantage in the selection rules; iii) move with constant in neutral neighbourhoods. 

According to many authors, organisms depend on their physical ability and accuracy to understand their neighbourhood \cite{Olfactory,perception, innatemite,Odour,MovementProfitable}; obstruction on the environmental scanning can also influence the making decision \cite{clues3}. Because of this, we sophisticate our model by simulating a wide range of scenarios, where decision making is not only based on external factors (the species densities in the individual's neighbourhood) but also on the organism's internal state\cite{doi:10.1002/ece3.4446,innatemite,SabelisII}:
i) individuals can perceive what is surrounding them in different ranges - the perception radius; ii) organisms may respond to the same sensory information with different strengths - the responsiveness.
Running many simulations and performing robust statistics, we quantify the effects of adaptive movement strategy on the spatial species densities. Furthermore, we calculate how the selection risk varies for a wide range of perception radii and adaptive responsiveness. 
By computing the spatial species densities and the selection risk, we make precise conclusions about the advantages and disadvantages that this evolutionary movement strategy brings at the population and individual level, respectively. 

The outline of this paper is as follows. In Sec.~\ref{sec2}, we describe the Methods, introducing the stochastic model, the adaptive movement strategy, the simulation implementation, and the model parameters. In Sec.~\ref{sec3}, we investigate how pattern formation is affected by the unevenness introduced by the adaptive movement strategy performed by organisms of species $1$. In Sec.~\ref{sec4}, we compute the species autocorrelation function and the characteristic length of the typical single-species spatial domain. We calculate the species densities in terms of
the responsiveness factor and perception radius in Sec.~\ref{sec5} and Sec. ~\ref{sec6}, respectively. We discuss the outcomes and present our conclusions in Sec.~\ref{sec7}.

%%%%%%%%%%%%%%%%%%%%%%%%%%%%%%%%%%%%%%%%%%%%%%%%%%%%%%%%%%%%%%%%%%%%%%%
\begin{figure*}
	\centering
    \begin{subfigure}{.19\textwidth}
        \centering
        \includegraphics[width=34mm]{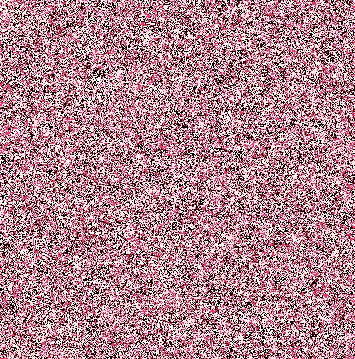}
        \caption{}\label{fig2a}
    \end{subfigure} %
   \begin{subfigure}{.19\textwidth}
        \centering
        \includegraphics[width=34mm]{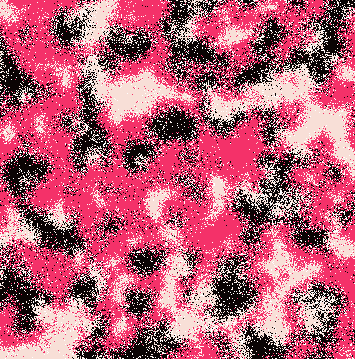}
        \caption{}\label{fig2b}
    \end{subfigure} 
            \begin{subfigure}{.19\textwidth}
        \centering
        \includegraphics[width=34mm]{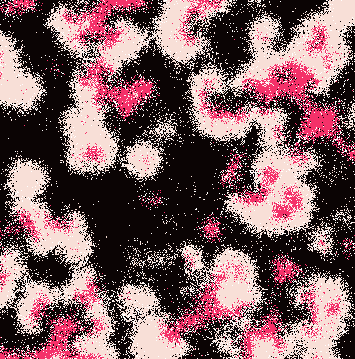}
        \caption{}\label{fig2c}
    \end{subfigure} 
           \begin{subfigure}{.19\textwidth}
        \centering
        \includegraphics[width=34mm]{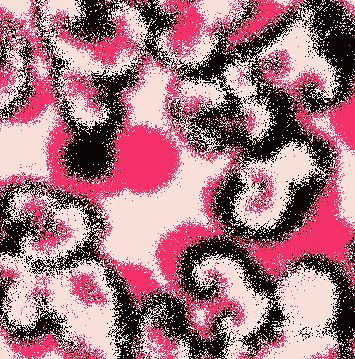}
        \caption{}\label{fig2d}
    \end{subfigure} 
   \begin{subfigure}{.19\textwidth}
        \centering
        \includegraphics[width=34mm]{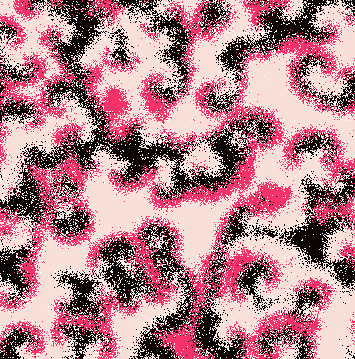}
        \caption{}\label{fig2e}
            \end{subfigure}
 \caption{Snapshots of a simulation starting from the spatial configuration in Fig.~\ref{fig2a}. The spatial configuration at $t=150$, $t=250$, $t=500$, and $t=4000$ are showed in Figs.~\ref{fig2b}, ~\ref{fig2c}, ~\ref{fig2d}, and ~\ref{fig2e}. The colours follow the scheme in Fig~\ref{fig1}; empty spaces appear as white dots. Video https://youtu.be/SY63KwkCjUw show the entire simulation.}
  \label{fig2}
\end{figure*}
%%%%%%%%%%%%%%%%%%%%%%%%%%%%%%%%%%%%%%%%%%%%%%%%%%%%%%%%%%%%%%%%%%%%%%%
%%%%%%%%%%%%%%%%%%%%%%%%%%%%%%%%%%%%%%%%%%%%%%%%%%%%%%%%%%%%%%%%%%%%%%%
\begin{figure}[t]
	\centering
        \begin{subfigure}{.5\textwidth}
        \centering
        \includegraphics[width=76mm]{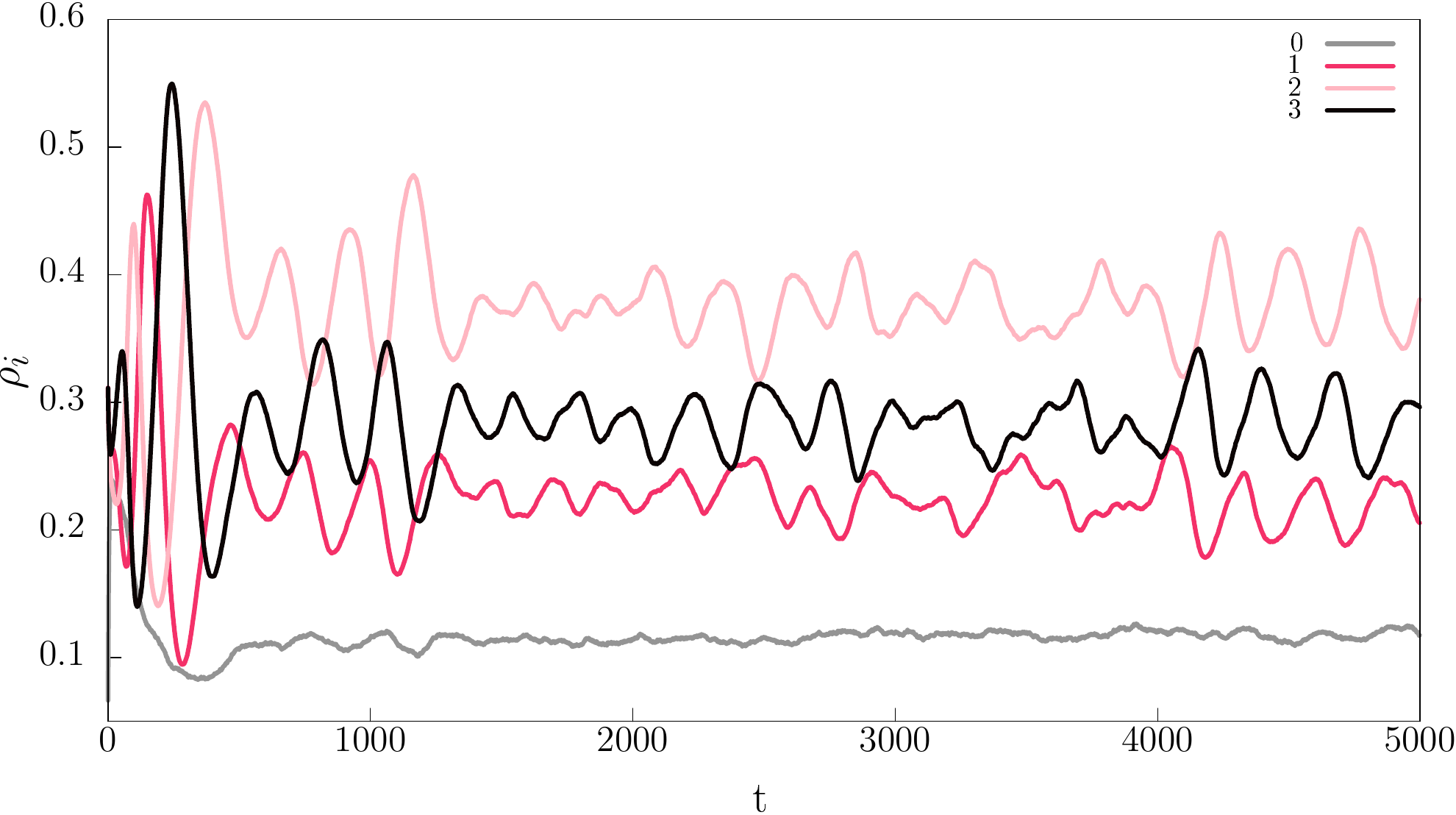}
        \caption{}\label{fig3a}
    \end{subfigure}\\
       \begin{subfigure}{.5\textwidth}
        \centering
        \includegraphics[width=76mm]{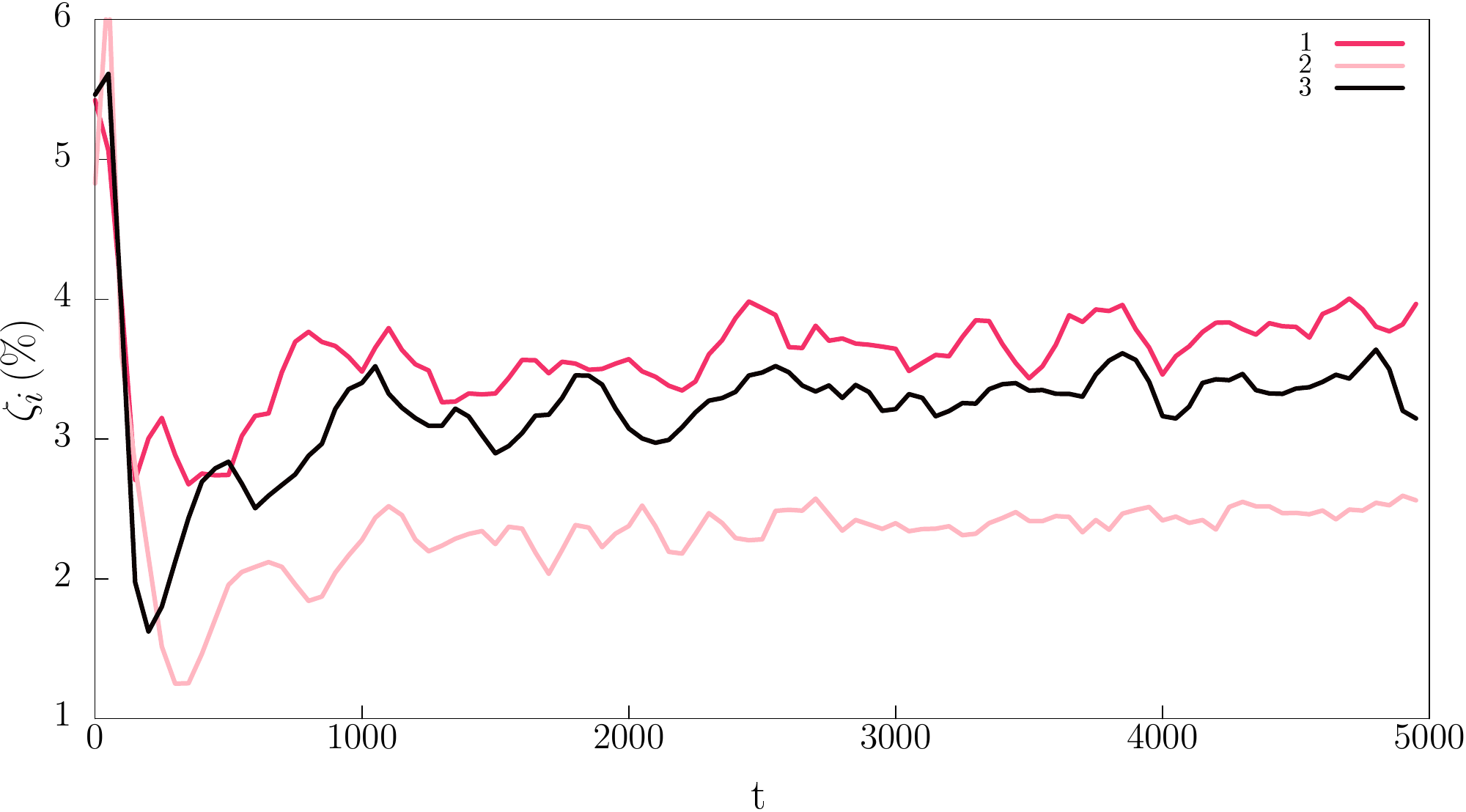}
        \caption{}\label{fig3b}
    \end{subfigure}
 \caption{
Dynamics of the species densities and selection risks for the simulation 
presented in Fig.~\ref{fig2}.
Figures~\ref{fig3a} and \ref{fig3b} depict $\rho_i$ and $\zeta_i$, respectively. The colours follow the scheme in Fig.~\ref{fig1}; grey lines depict the dynamics of the density of empty spaces. The error bars show the standard deviation.}
  \label{fig3}
\end{figure}
%%%%%%%%%%%%%%%%%%%%%%%%%%%%%%%%%%%%%%%%%%%%%%%%%%%%%%%%%%%%%%%%%%%%%%%

%%%%%%%%%%%

\section{Methods}
\label{sec2}

%%%%%%%%%%%%
\subsection{The model}
%%%%%%%%%%%%%%%%%%%%%%%%%%%%%%%%%%%%%%%%%%%%%%%%%%%%%%%%%%%%%%%%%%%%%%%

We performed agent-based stochastic simulations of a cyclic nonhierarchical system composed of $3$ species, whose dominance is described by the rock-paper-scissors game rules. In our model, individuals of one out of the species can control their mobility rate according to the local species densities. The adaptive movement strategy allows an individual to judge if it should stay or go away from its place. Our numerical implementation follows a standard algorithm widely employed in studies of spatial biological systems \cite{Reichenbach-N-448-1046,uneven,Szolnoki_2020}. The dynamics of individuals' spatial organisation was simulated in square lattices with periodic boundary conditions, following the rules: selection, reproduction, and mobility. We assumed the May-Leonard numerical implementation, where it is not assumed a conservation law for the total number of individuals \cite{leonard}. As each grid point contains at most one individual, the maximum number of individuals is $\mathcal{N}$, the total number of grid points.

Figure~\ref{fig1} illustrates the main rules of our simulations, with dark pink, light pink, and black representing species $1$, $2$, and $3$, respectively. The arrows show the cyclic dominance of the selection interactions: organisms of species $i$ beat individuals of species $i+1$, with $i=1,2,3$, with the cyclic identification $i=i\pm3\,k$, where $k$ is an integer. The purple bars illustrate the mobility interactions among organisms of every species; the open circles represent the adaptative movement tactic: individuals of species $1$ are attracted by species $2$ (the attractant species) and repelled by species $3$ (the repellent species).

The initial conditions were prepared so that the number of individuals is the same for all species, i.e., $I_i\,=\,\mathcal{N}/3$, with $i=1,2,3$. The initial spatial configuration was built by randomly distributing one individual at a random grid point. At each timestep, the individuals' displacement is altered by the implementation of one of the spatial interactions:
\begin{itemize}
\item 
Selection: $ i\ j \to i\ \otimes\,$, with $ j = i+1$, where $\otimes$ means an empty space: whenever one selection interaction occurs, the grid point occupied by the individual of species $i+1$ becomes an empty space.
\item
Reproduction: $ i\ \otimes \to i\ i\,$: when one reproduction is realised, a new organism of species $i$ occupies an empty space.
\item 
Mobility: $ i\ \odot \to \odot\ i\,$, where $\odot$ means either an individual of any species or an empty site: an individual of species $i$ switches positions with another individual or with an empty space.
\end{itemize}

%%%%%%%%%%%%
\subsection{Simulations}
%%%%%%%%%%%%

We work with the Moore neighbourhood, i.e., individuals may interact with one of their eight nearest neighbours. The simulation algorithm follows three steps: i) randomly selecting an active individual; ii) raffling one interaction to be executed; iii) drawing one of the four nearest neighbours to suffer the sorted interaction. If the interaction is realised, one timestep is counted. Otherwise, the three steps are redone. The time necessary to $\mathcal{N}$ timesteps to occur is one generation, our time unit.

In our simulations, the interactions are implemented stochastically, with probabilities computed according to the selection, reproduction, and mobility rates. All rates are constant except for the mobility of individuals of species $1$, whose proportion of the grid area explored per unit of time varies according to the local attractiveness \cite{Reichenbach-N-448-1046}.
For every species, selection and reproduction interactions are implemented following the probabilities $s$ and $r$, respectively. These probabilities are not dependent on the space, being the same for every individual in the grid. Although the maximum probability is $m$ for all species, $s+r+m=1$, we assume that:
i) for all organisms of species $2$ and $3$, the mobility probability is the constant, $m_2\,=\,m_3\,=\,m/2$; ii) for individuals of species $1$, the mobility probability, $m_1$, depends on the local species densities so that, $m_1\,<\,m/2$ in attractive zones and $m_1\,>\,m/2$ in hostile regions.

\subsection{Implementation of the adaptive movement}

The numerical implementation of the adaptive movement behaviour of organisms of species $1$ follows the step: 
\begin{enumerate}
\item
We define the maximum distance one individual can perceive a neighbour:
the perception radius $R$, an integer parameter, measured in lattice spacing.
\item
The grid points within the disk of radius $R$ around the active individual are examined: it is computed the local densities of individuals of species $2$ ($d_2$) and species $3$ ($d_3$).
\item
We calculate the local attractiveness, $\mathcal{A}$ as the difference between the local densities of individuals of species $2$ and $3$:
\begin{equation}
\mathcal{A} = d_2\,-\,d_3.
\end{equation}
\item 
We define the local adaptability function $\mathcal{F}=\mathcal{F}(d_2,d_3)$, a real function given by 
\begin{equation}
\mathcal{F}\,=\,\frac{1}2\,\Big[1\,+\,\tanh \left(-\epsilon\,\mathcal{A}\right)\Big],
\end{equation}
where $\epsilon$ is a real parameter that defines the organism's responsiveness to the local stimuli; the local density of organisms of species $1$ does not influence the organism's mobility. 
\item
Finally, the active individual's mobility probability is calibrated within the range $0<m_1<m$, according to 
\begin{equation}
m_1\,=\,m\,\mathcal{F}.
\end{equation}
\end{enumerate}

Figure~\ref{fig1b} shows how the local attractiveness function $\mathcal{A}$ controls the effective mobility probability of organisms
of species $1$, for $m=0.8$. The grey, brown, blue, green, yellow, orange, and purple lines indicate the organisms' responsiveness for $\epsilon=0.0$ (standard model), $\epsilon=0.5$, $\epsilon=1.0$, $\epsilon=1.5$, $\epsilon=2.0$, $\epsilon=3.0$, and $\epsilon=4.0$, respectively. 
The illustration shows that: i) individuals' reaction to adjust their speed is more accentuated as $\epsilon$ increases; ii) the minimum effective mobility probability is $[1.0+\tanh(-\epsilon)]/2$, occurring if the organism's neighbourhood is composed exclusively by organisms of species $2$ (attractant) - for example, for $\epsilon=2.0$, one has $m_1=0.018$; iii) the maximum effective mobility probability is $[1.0+\tanh(\epsilon)]/2$, in zones where $d_3=1$ (repellent) - for example, the maximum mobility probability is $m_1=0.982$, for 
$\epsilon=2.0$.

%%%%%%%%%%%%%%%%%%%%%%%%%%%
\subsection{Spatial Patterns}
%%%%%%%%%%%%%%%%%%%%%%%%%%%
\begin{figure*}
	\centering
	\centering
    \begin{subfigure}{.18\textwidth}
        \centering
        \includegraphics[width=32mm]{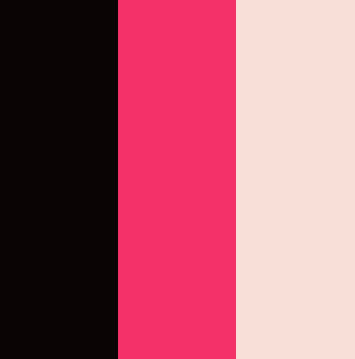}
        \caption{}\label{fig4a}
    \end{subfigure} %
   \begin{subfigure}{.18\textwidth}
        \centering
        \includegraphics[width=32mm]{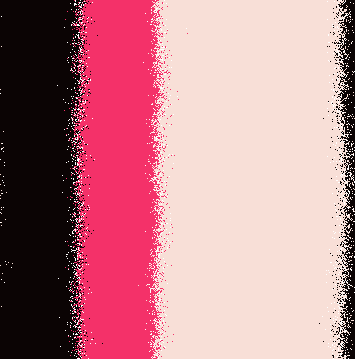}
        \caption{}\label{fig4b}
    \end{subfigure} 
            \begin{subfigure}{.18\textwidth}
        \centering
        \includegraphics[width=32mm]{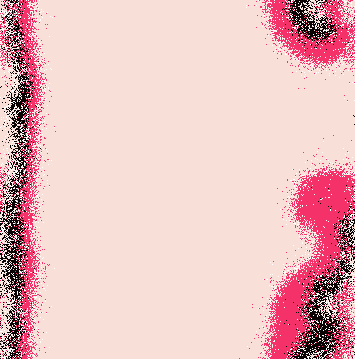}
        \caption{}\label{fig4c}
    \end{subfigure} 
           \begin{subfigure}{.18\textwidth}
        \centering
        \includegraphics[width=32mm]{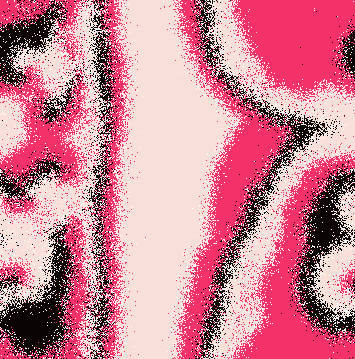}
        \caption{}\label{fig4d}
    \end{subfigure} 
   \begin{subfigure}{.18\textwidth}
        \centering
        \includegraphics[width=32mm]{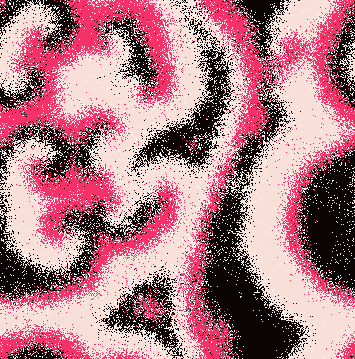}
        \caption{}\label{fig4e}
            \end{subfigure}\\
   \begin{subfigure}{.18\textwidth}
        \centering
        \includegraphics[width=32mm]{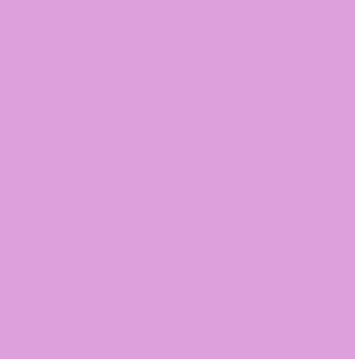}
        \caption{}\label{fig4f}
    \end{subfigure} 
       \begin{subfigure}{.18\textwidth}
        \centering
        \includegraphics[width=32mm]{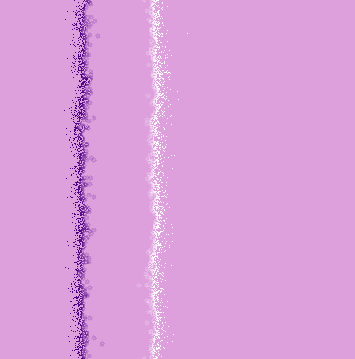}
        \caption{}\label{fig4g}
    \end{subfigure}
          \begin{subfigure}{.18\textwidth}
        \centering
        \includegraphics[width=32mm]{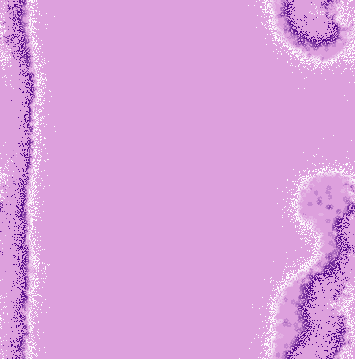}
        \caption{}\label{fig4h}
    \end{subfigure} 
       \begin{subfigure}{.18\textwidth}
        \centering
        \includegraphics[width=32mm]{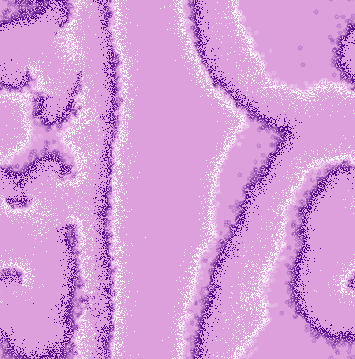}
        \caption{}\label{fig4i}
    \end{subfigure} 
      \begin{subfigure}{.18\textwidth}
        \centering
        \includegraphics[width=32mm]{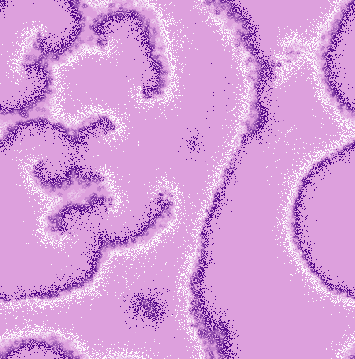}
        \caption{}\label{fig4j}
    \end{subfigure} 
 \caption{Snapshots of a simulation starting from the spatial configuration in Fig.~\ref{fig4a}. The spatial configuration at $t=1000$, $t=3200$, $t=3600$, and $t=4200$ are showed in Figs.~\ref{fig4b}, ~\ref{fig4c}, ~\ref{fig4d}, and ~\ref{fig4e}, respectively. The respective local variation of the mobility probability is depicted in Figs.~\ref{fig4f}, ~\ref{fig4g}, ~\ref{fig4h}, ~\ref{fig4i}, and ~\ref{fig4j}. Videos https://youtu.be/VIjearUHAN4 and https://youtu.be/sOeGdnU-Ghk show 
the dynamics of the spatial configuration and local mobility during the entire simulation for the species segregation.}
  \label{fig4}
\end{figure*}
%%%%%%%%%%%%%%%%%%%%%%%%%%%%%%%%%%%%%%%%%%%%%%%%%%%%%%%%%%%%%%%%%%%%%%%

We first performed a single simulation for observing the spatial patterns.
The realisations were performed in square lattices with $500^2$ grid points for a timespan of $5000$ generations. The perception radius was set $R=2$, while $s\,=\,r\,=\,0.1$ and $m = 0.8$. 
We captured $500$ snapshots of the lattice (in intervals of $10$ generations); then, we used the snapshots to produce the video https://youtu.be/SY63KwkCjUw. Figures \ref{fig2a}, ~\ref{fig2b}, ~\ref{fig2c}, and \ref{fig2d} show snapshots captured from the simulations at $t=0$ (initial conditions), $t=150$, $t=250$, $t=500$, and $t=4000$ generations - the snapshots were chosen to highlight the pattern formation process. Organisms of species $1$, $2$, and $3$, are showed by dark pink, light pink, and black dots, respectively; empty spaces are represented by white dots. 

To investigate the pattern formation in more detail, we prepared a single simulation starting from the particular initial condition shown in Fig.~\ref{fig3a}. Each species occupies a third of the grid with all individuals starting with mobility probability $m_1\,=\,m_2\,=m_3\,=\,0.4$. Beyond taking the snapshots of the species configurations, we depict the spatial distribution of the local mobility probability - descending purple colours show the spatial variation of mobility probability, with lighter and darker purple representing the lower and higher mobility probabilities.
The results are depicted in Fig.~\ref{fig3} and videos https://youtu.be/VIjearUHAN4 and https://youtu.be/sOeGdnU-Ghk.

As soon as the simulation starts, the width of the rings occupied by each species changes linearly in time. Considering that the area occupied by species $i$ is defined by the total number of organisms of species $i$, one has $$
\dot{\delta_i}\,=\,\frac{\dot{I_{i}}}{\sqrt{\mathcal{N}}},$$
with $i=1,2,3$; the dot stands for the time derivative, $\delta_i$ is the width of the ring of species $i$, and $\sqrt{\mathcal{N}}$ is the torus cross-section perimeter.
%%%%%%%%%%%%%%%%%%%%%%%%%%%%%%%%%%%%%%%%%%%%%%%%%%%%%%%%%%%%%%%%%%%%%%%
\subsection{Autocorrelation function}
%%%%%%%%%%%%%%%%%%%%%%%%%%%%%%%%%%%%%%%%%%%%%%%%%%%%%%%%%%%%%%%%%%%%%%%

We calculate the spatial autocorrelation function $C_i(r)$, with $i=1,2,3$, in terms of radial coordinate $r$, where $r=|\vec{r}|=x+y$ is the Manhatan distance between $(x,y)$ and $(0,0)$. For this purpose, we define the function $\phi_i(\vec{r})$ that represents the presence of an organism of species $i$ in the position $\vec{r}$ in the lattice \cite{Moura,MENEZES2022101606,auto1,auto2}.
Using the mean value $\langle\phi_i\rangle$, we compute the Fourier transform
\begin{equation}
\varphi_i(\vec{\kappa}) = \mathcal{F}\,\{\phi_i(\vec{r})-\langle\phi_i\rangle\}.
\end{equation}
Subsequently, we find the spectral densities
\begin{equation}
S_i(\vec{k}) = \sum_{k_x, k_y}\,\varphi_i(\vec{\kappa}).
\end{equation}

Therefore, the autocorrelation function is given by the normalised inverse Fourier transform
\begin{equation}
C_i(\vec{r}') = \frac{\mathcal{F}^{-1}\{S_i(\vec{k})\}}{C(0)},
\end{equation}
which we write as 
a function of the radial coordinate $r$ as
\begin{equation}
C_i(r') = \sum_{|\vec{r}'|=x+y} \frac{C_i(\vec{r}')}{min\left[2N-(x+y+1), (x+y+1)\right]}.
\end{equation}
The typical size of the spatial domain occupied by organisms of species $i$ is found by employing 
threshold $C_i(l_i)=0.15$, where $l_i$ is the characteristic length scale for spatial domains of species $i$.

The autocorrelation function was found by averaging the results obtained by $100$ simulations using lattices with $300^2$ grid points, with each simulation starting from different random initial conditions. We used the parameters $s\,=\,r\,=\,0.1$, $m= 0.8$, $R=2$. 
We computed the mean autocorrelation function $C_i(r)$ in the radial coordinate $r$ and calculated the characteristic length for each species employing the spatial configuration at $5000$. The simulations were repeated for various values of $\epsilon$.
Figure~\ref{fig4} shows the autocorrelation function as a function of the radial coordinate $r$ for the standard model and the adaptive movement strategy for various values of $\epsilon$. The colours follow the scheme in Fig.~\ref{fig1}, where dark pink, light pink, and black circles indicate the mean values for the species $1$, $2$, and $3$, respectively. Grey circles show the results for $\epsilon=0$, where the mean values are the same for all species. The standard deviation is shown by error bars. The horizontal black line indicates the threshold used to calculate the characteristic length, $C(l_i)\, =\, 0.15$. The inset figure show $l_i$ for $0\,\leq\,\epsilon\,\leq4$.

%%%%%%%%%%%%%%%%%%%%%%%%%%%%%%%%%%%%%%%%%%%%%%%%%%%%%%%%%%%%%%%%%%%%%%%
\subsection{Species densities and selection risk}
%%%%%%%%%%%%%%%%%%%%%%%%%%%%%%%%%%%%%%%%%%%%%%%%%%%%%%%%%%%%%%%%%%%%%%%

%%%%%%%%%%%%%%%%%%%%%%%%%%%%%%%%%%%%%%%%%%%%%%%%%%%%%%%%%%%%%%%%%%%%%%%
To understand how the adaptive movement behaviour affects species population dynamics, we defined the species spatial densities $\rho_i$, with $i=1,2,3$, as the fraction of the grid occupied by individuals of species $i$ at time $t$: $\rho_i\,=\,I_i/\mathcal{N}$. The proportion of empty space is denoted by $\rho_0$. Moreover, we studied the influence of adaptive movement tactic on the selection risk of individuals of species $i$, $\zeta_i$. In this sense, we used the following algorithm: i) the total number of individuals of species $i$, is counted at the beginning of each generation; ii) it is computed the number of individuals of species $i$ caught during the generation; iii) it is calculated $\zeta_i$ as the ratio between the number of selected individuals and the initial amount.

Figures ~\ref{fig3a}, and ~\ref{fig3b} show $\rho_i$ and $\zeta_i\,(\%)$ as functions of the time for the simulations presented in Fig. ~\ref{fig2}.
The colour follows the scheme in Fig.~\ref{fig1}. The outcomes presented in Figure \ref{fig3b} were averaged for every $50$ generation.

Subsequently, we run groups of $100$ simulations in grids of $300^2$ sites, running until $5000$ generations to further explore how the parameters $\epsilon$, and $R$ influence the adaptive movement strategy in cyclic models. 
We computed the mean value of the spatial species densities and selection risks from simulations starting from different initial conditions, for $s\,=\,r\,=\,0.1$ $m\,=\,0.8$. Aiming to avoid the density fluctuations inherent in the pattern formation process, we averaged the results considering the numerical data of the second half of the simulations.
The collections of numerical experiments were organised as follows.
\begin{itemize}
\item
Fixing $R\,=\,2$, we run simulations for the range of responsiveness factor $0\,\leq\,\epsilon\,\leq\,4$, with $\Delta\,\epsilon\,=\,0.5$. 
\item
Assuming $\epsilon\,=\,2.0$, the realisations were performed for the perception radii $R\,=\,1,2,3,4$. 
\end{itemize}
The outcomes appear in Figs.~\ref{fig6a}, ~\ref{fig6b} (species densities), ~\ref{fig7a}, ~\ref{fig7b} (selection risk), where circles represent the mean value, whereas error bars indicate the standard deviation. 
 
%%%%%%%%%%%%%%%%%%%%%%%%%%%%%%%%%%%%%%%%%%%%%%%%%%%%%%%%%%%%%%%%%%%%%%%
\begin{figure}[t]
	\centering
	\includegraphics[width=90mm]{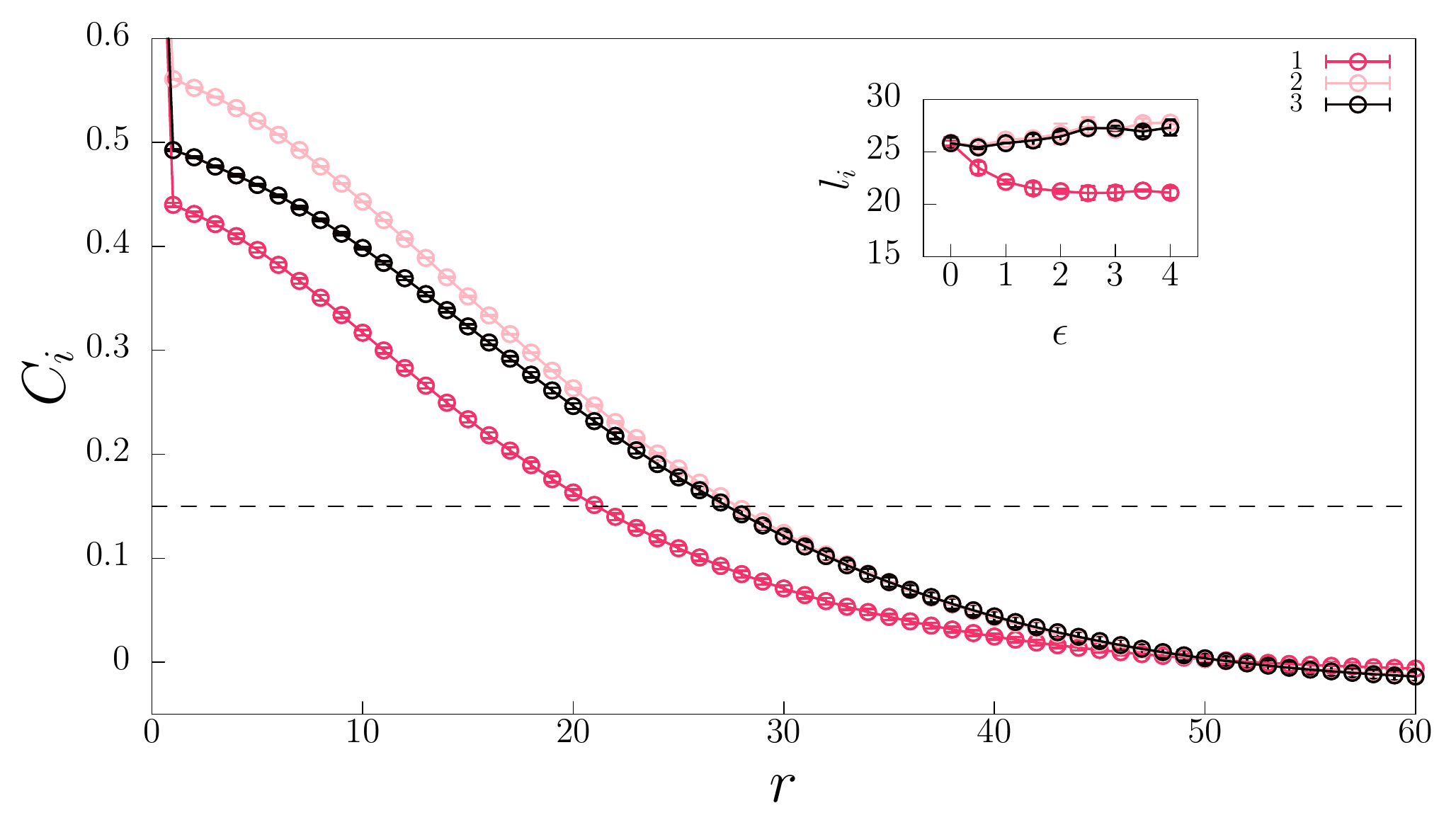}
    \caption{Autocorrelation functions $C_i$ and characteristic length $l_i$. The colours follow the scheme in Fig.~\ref{fig1}. The error bars indicate the standard deviation. The horizontal dashed black line indicates the threshold assumed to calculate the characteristic length, depicted in the inset for a range of $\epsilon$.}
  \label{fig5}
\end{figure}
%%%%%%%%%%%%%%%%%%%%%%%%%%%%%%%%%%%%%%%%%%%%%%%%%%%%%%%%%%%%%%%%%%%%%%%

%%%%%%%%%%%%%%%%%%%
\section{Spatial patterns}
\label{sec3}
%%%%%%%%%%%%%%%%%%
%%%%%%%%%%%
Figure~\ref{fig2} shows snapshots captured from a single simulation
with adaptive movement strategy of organisms of species $1$. 
Organisms of species $1$, $2$, and $3$ are shown by dark pink, light pink, and black dots, respectively; white dots indicate empty spaces. Since the realisations started from random initial conditions (Fig.~\ref{fig2a}), selection interactions are frequent in the initial stage of the simulations. The unevenness introduced by the clever movement of individuals of species $1$ leads to an alternate territorial dominance among species in the initial transient stage of pattern formation (\cite{uneven}). For example, Figs.~\ref{fig2b} and ~\ref{fig2c} show the territorial control of species $1$ being conquered by species $3$ - this happens because of the rules of the cyclic spatial rock-paper-scissors game. Subsequently, spiral waves are formed, with organisms of a single species inhabiting departed spatial domains.
The irregular spiral formation (Fig.~\ref{fig2d} and Fig.~\ref{fig2e}) reflects the turbulence on the spatial segregation of species when individuals of species $1$ no longer move with constant mobility probability but scan the environment to produce the appropriate motor response. See also video https://youtu.be/SY63KwkCjUw.

To understand the causes of the irregular spiral formation caused by the adaptive movement tactic of organisms of species $1$, we run a simulation starting from the prepared initial conditions in Fig.~\ref{fig4a}. Because of the periodic boundary conditions, the individuals are initially constrained to live in single-species rings - each species fills one-third of the grid. Figure ~\ref{fig4f} 
shows that the initial mobility probability for every species is $m\,_i\,=\,0.4$, with $i=1,2,3$, as indicated by the purple colour. As soon as the simulation starts, selection interactions provoke the movement of the rings from left to right. However, while all individuals of species $2$ (light pink) and $3$ (black) continue moving with constant mobility probability, organisms of species $1$ (dark pink) located in the nearby of attractant or repellent species adjust their mobility probability according to the local reality. This happens because individuals of species $1$ on the ring front boundary sense the presence of individuals of species $2$; analysing their neighbourhood, they conclude they are in a comfort zone, and consequently, they diminish their mobility probability - as depicted by light purple dots in Fig.~\ref{fig4g} ($m \approx 0$). According to Fig.~\ref{fig4b}, the consequence is a reduction of the speed of the dark pink domain that causes a widening of the light pink domain and a narrowing of the dark pink region. In addition, individuals of the species $1$ close to the black ring perceive the species $3$; this indicates to escape increasing their mobility probability, as represented by the dark purple dots in Fig.~\ref{fig4g} ($m \approx 0.8$). The result is that the black ring also narrows because individuals of species $1$ can move away from individuals of species $3$ faster than individuals of species $3$ advance on them. We calculated the time variation of each ring width by quantifying how the total number of individuals changes in time: $\delta_1\,=\,-0.0279$, $\delta_2\,=\,0.0412$, and $\delta_3\,=\,-\,0.01609$ grid points per generation.

As time passes, the narrowing of the dark pink and black rings accentuates until a transition happens: the width of the black domain becomes too short that individuals of species $1$ manage to cross it reaching the light pink domain, as shown in Fig.~\ref{fig4c} and Fig.~\ref{fig4h}. The abundance of individuals of species $2$ allows the proliferation of species $1$, followed by a growth of the population of species $3$. This generates spiral waves that spread on the light pink ring, as appears in Fig.~\ref{fig4d}. The main point is that the organisms in the spiral arm of species $1$ move with different velocities: individuals in the front have lower mobility probabilities than organisms in the back, as shown in Fig.~\ref{fig4i}. Consequently, the spiral arm of individuals of species $1$ (dark pink) is narrower than both other arms, as shown in Fig.~\ref{fig4e} and Fig.~\ref{fig4j}. Videos https://youtu.be/VIjearUHAN4 and https://youtu.be/sOeGdnU-Ghk show the dynamics of the individuals' spatial distribution and the mobility probability for the entire simulation.

Figure~\ref{fig3a} depicts the species densities and selection risks in the simulations presented in Fig.~\ref{fig2}.
The grey colour depicts the density of empty space ($\rho_0$) while dark pink, light pink, and black show the results for species $1$, $2$, and $3$, respectively. The outcomes reveal that the adaptive movement tactic is disadvantageous for species $1$ in terms of territorial dominance. 
The outcomes also indicate that the discrepancy in the species densities increases with the responsiveness factor since the individuals can react more readily to the sensory information. 
Figure~\ref{fig3b} shows how the risk of an organism being eliminated in the cyclic game of simulation in Fig.~\ref{fig2}. Our outcomes show that 
the selection risk of species $1$ and $3$ increases when individuals of species $1$ move strategically. Conversely, an individual of species $2$ sees its risk of being caught diminishes. 
%%%%%%%%%%%%%%%%%%%%%%%%%%%%%%%%%%%%%%%%%%%%%%%%%%%%%%%%%%%%%%%%%%%%%

%%%%%%%%%%%%%%%%%%%%%%%%%%%
\section{Characteristic Length of the Single-Species Domains}
\label{sec4}
%%%%%%%%%%%%%%%%%%%%%%%%%%%

Figure~\ref{fig2} revealed that the spiral patterns are deformed whether organisms of species $1$ adjust their velocity following environmental clues. Notably, the average agglomeration size is smaller for organisms of species $1$. Therefore, we now aim to quantify the characteristic length for domains formed of each species. For this purpose, we first computed the spatial autocorrelation function $C_i(r)$, with $i=1,...,3$, whose results are depicted in Fig.~\ref{fig4}. Dark pink, light pink, and black depict the outcomes for species $1$, $2$, and $3$, respectively.
Observing the results obtained for $\epsilon=2.0$, we confirmed that organisms of species $1$ are less spatially correlated when compared with individuals of other species. We thereby measured the characteristic length $l_i$, defined as $C_{i}(l_{i})=0.15$ (grey line in Fig.~\ref{fig4}) for a range of $\epsilon$. The results are presented in the inset of Fig.~\ref{fig4} for $0\,\leq\,\epsilon\,\leq\,4$.
As the responsiveness factor grows, the effect on the average size of the single-species spatial domains becomes more relevant: the characteristic length for areas controlled by organisms of species $1$
nonlinearly decreases, while other domains' characteristic lengths elongate, approximately the same for species $1$ and $2$.

%%%%%%%%%%%%%%%%%%%%%%%%%%%
\section{Species Densities}
\label{sec5}
\begin{figure}[t]
	\centering
    \begin{subfigure}{.51\textwidth}
    \centering
    \includegraphics[width=79mm]{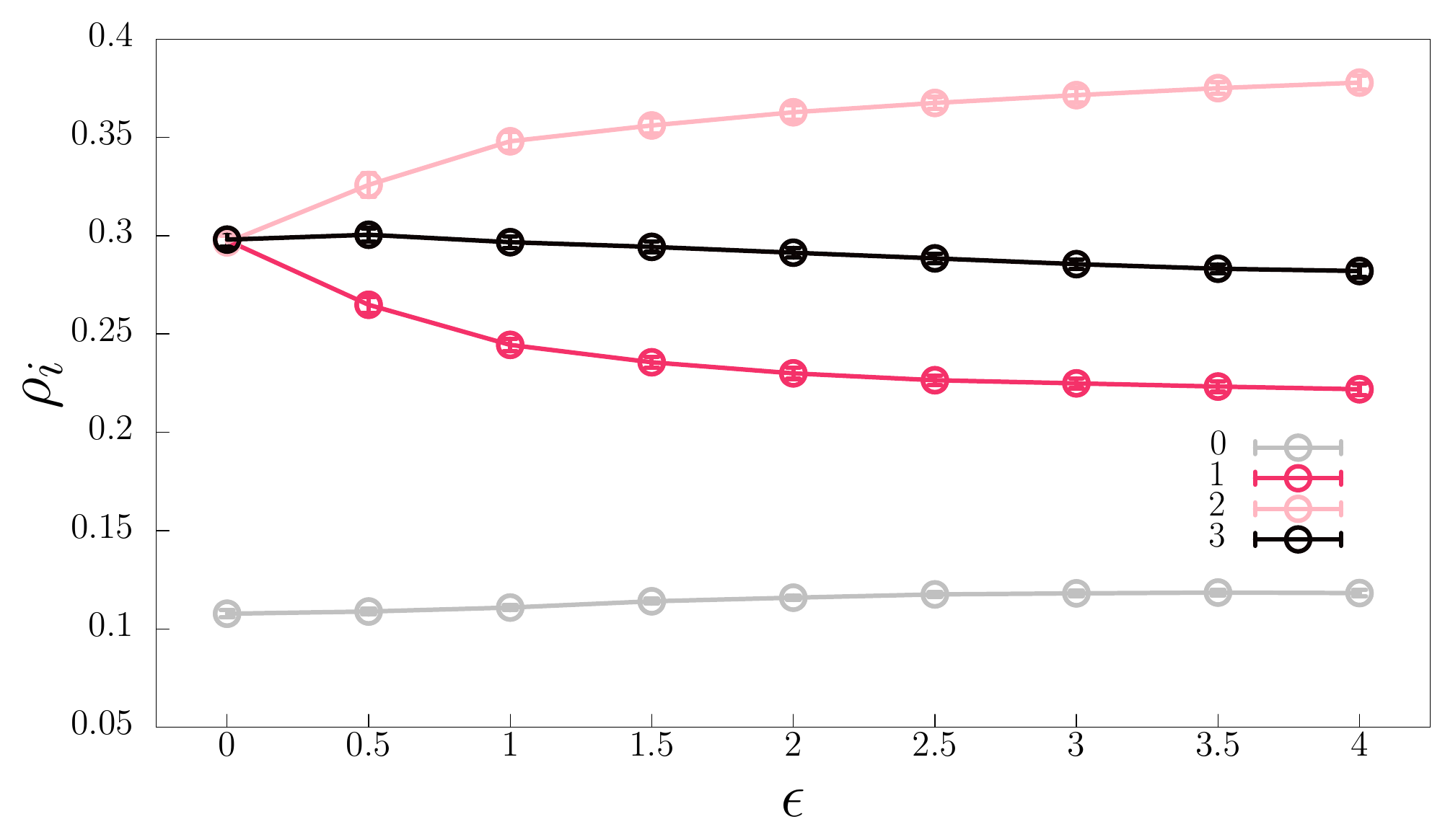}
    \caption{}\label{fig6a}
  \end{subfigure}\\
      \begin{subfigure}{.51\textwidth}
    \centering
    \includegraphics[width=79mm]{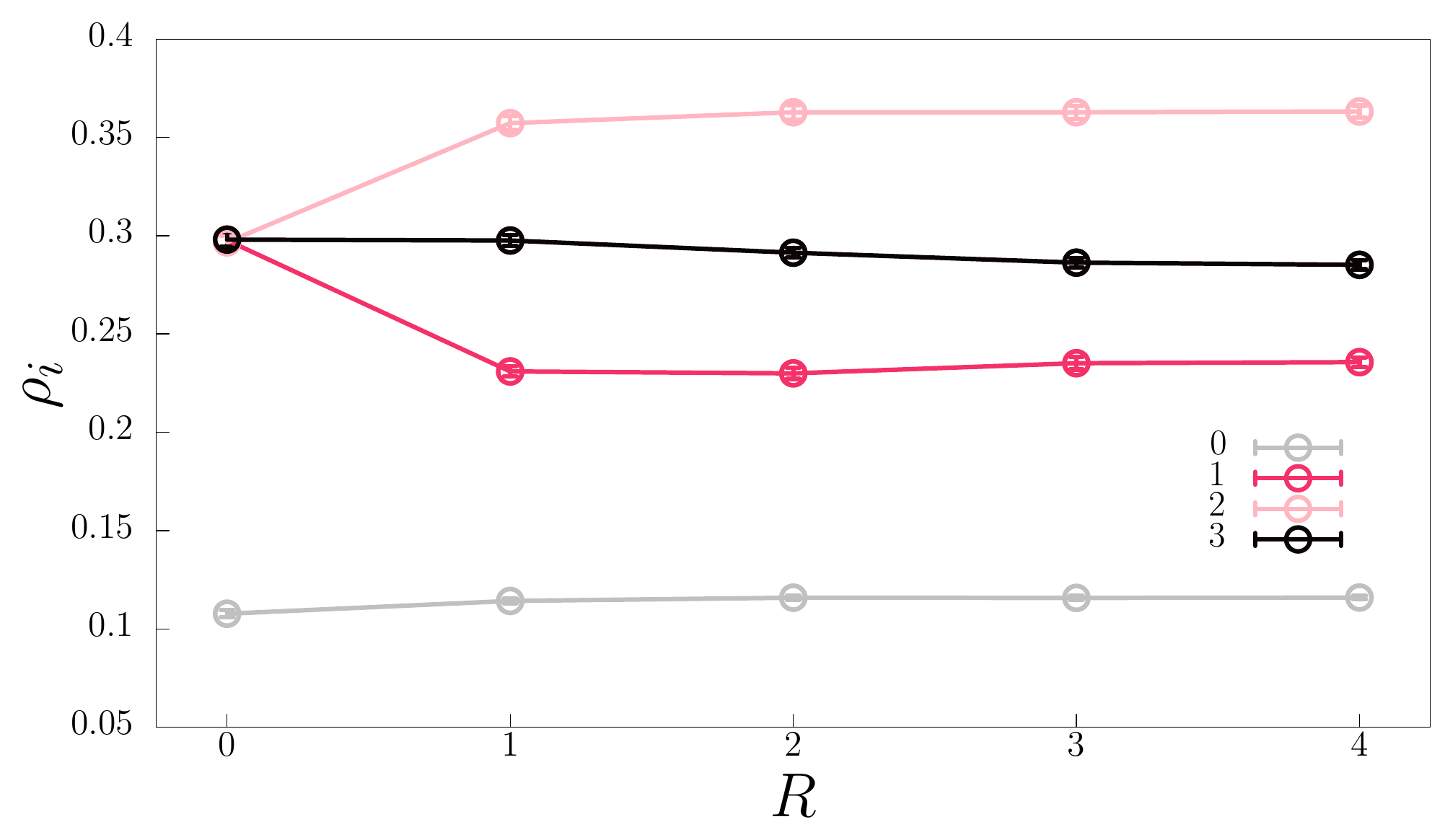}
    \caption{}\label{fig6b}
  \end{subfigure}
    \caption{Mean species densities resulting from the adaptive movement strategy of species $1$. Figures~\ref{fig6a} and \ref{fig6b} show the outcomes 
in terms of $\epsilon$ and $R$, respectively. The error bars show the standard deviation. The colours follow the scheme in Fig.~\ref{fig1}.}
  \label{fig6}
\end{figure}

%%%%%%%%%%%%%%%%%%%%%%%%%%%
Now, we explore the effects of the adaptive movement strategy on the population dynamics by investigating the role of the responsiveness factor and the perception radius in the species densities. 
Figures ~\ref{fig6a},  and ~\ref{fig6b} show how species densities vary in terms of $\epsilon$ and $R$, respectively. The mean value was computed by averaging the outcomes from sets of $100$ simulations; the error bars represent the standard deviation. Overall, our discoveries unveiled that the more readily and accurately organisms respond to the local stimulus, the more species populations are affected. 
As $\epsilon$ grows, the territorial dominance of species $2$ is enhanced, with negative variation for the populations of species $1$ and $3$. However, we found that the mean species densities do not vary significantly with $R$. Moreover, for $R>1$, there is a slight augment of the portion of the grid occupied by species $1$ in detriment of the total area controlled by species $3$.

%%%%%%%%%%%%%%%%%%%%%%%%%%%%%%%%%%%%%%%%%%%%%%%%%%%%%%%%%%%%%%%%%%%%%%%
%%%%%%%%%%%%%%%%%%%%%%%%%%%%%%%%%%%%%%%%%%%%%%%%%%%%%%%%%%%%%%%%%%%%%%%

%%%%%%%%%%%%%%%%%%%%%%%%%%%%%%%%%%%%%%%%%%%%%%%%%%%%%%%%%%%%%%%%%%%%%%%
\begin{figure}[t]
	\centering
    \begin{subfigure}{.51\textwidth}
    \centering
    \includegraphics[width=79mm]{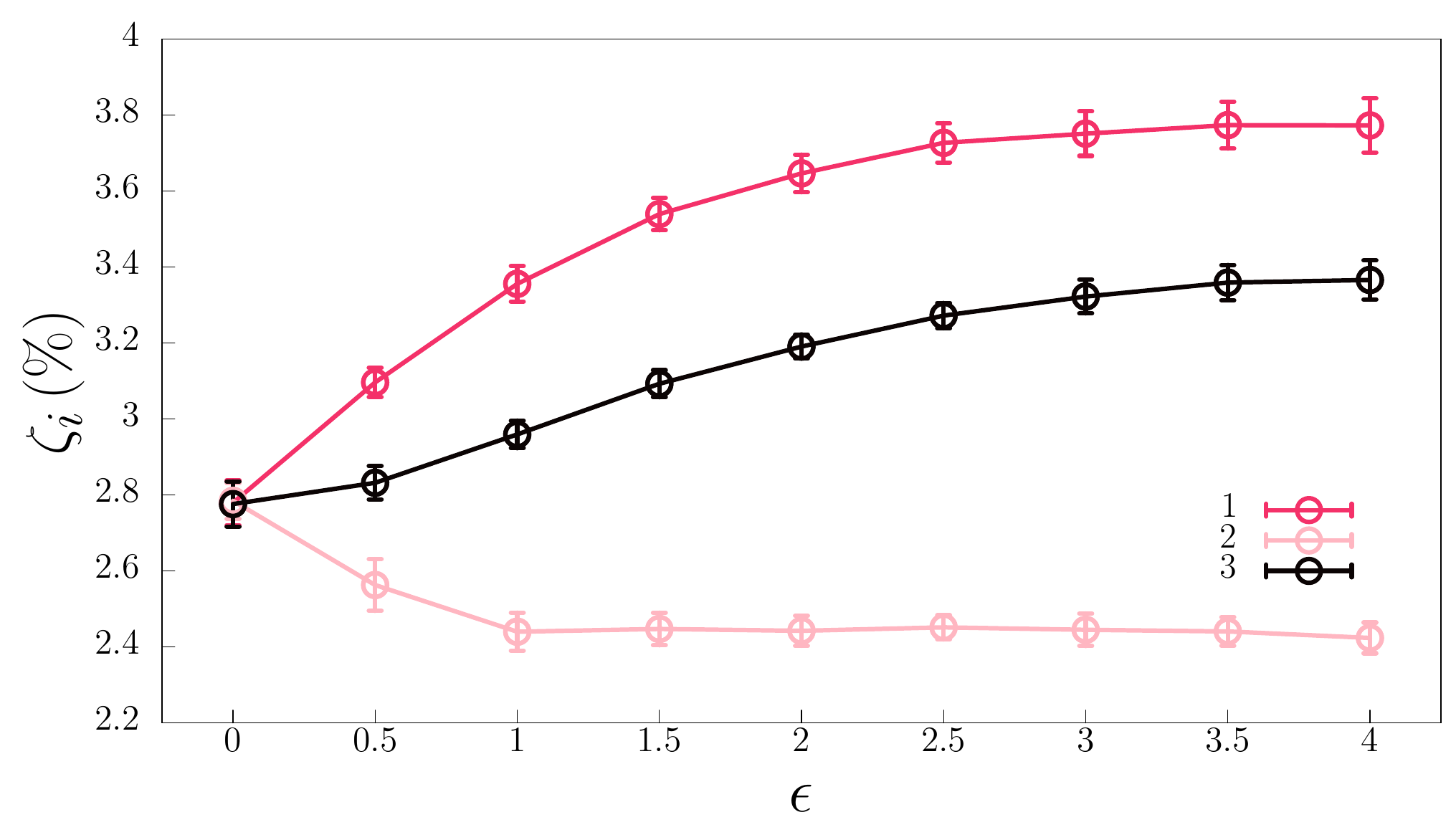}
    \caption{}\label{fig7a}
  \end{subfigure}\\
      \begin{subfigure}{.51\textwidth}
    \centering
    \includegraphics[width=79mm]{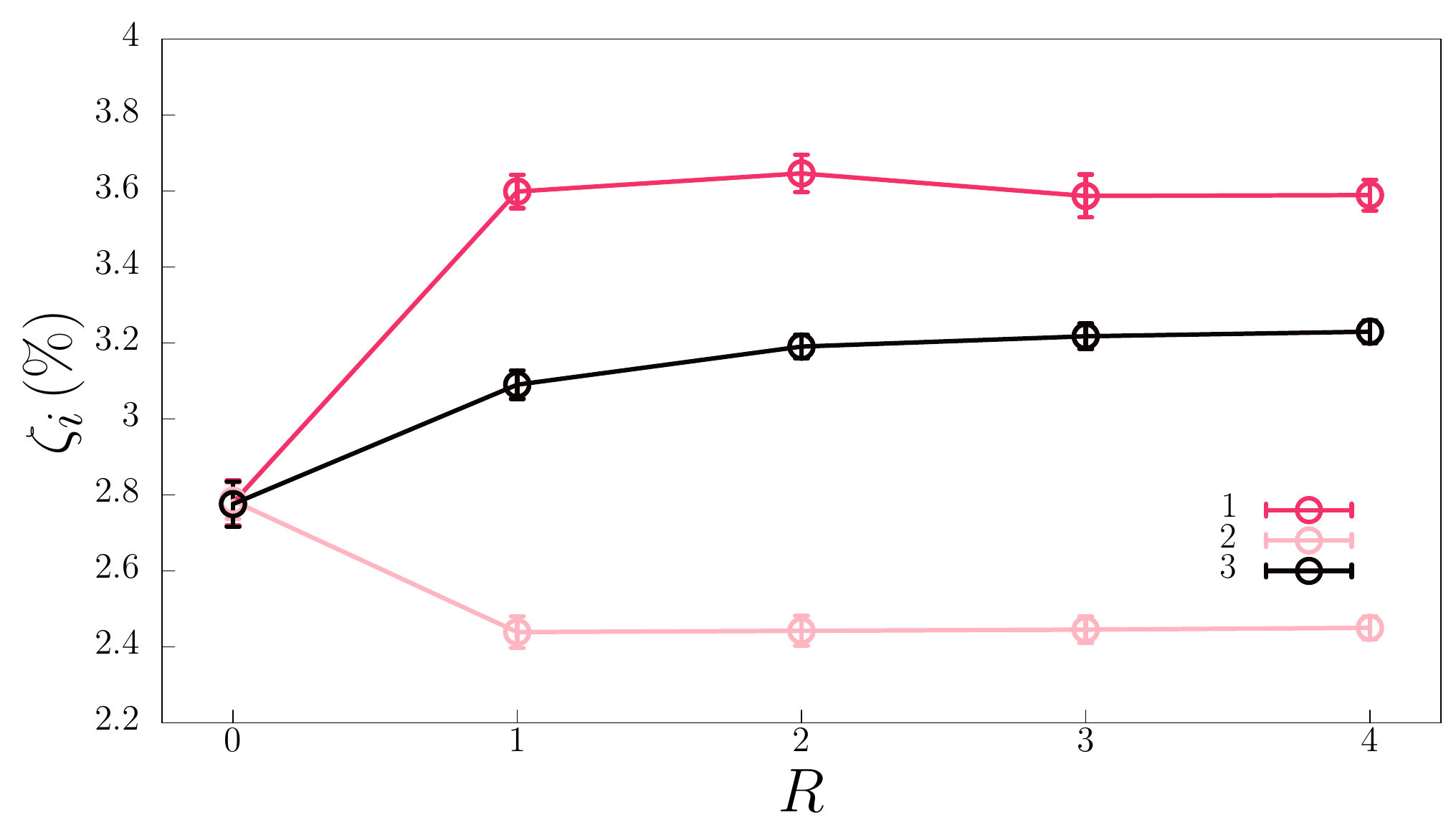}
    \caption{}\label{fig7b}
  \end{subfigure}
    \caption{Changes in the species selection risk due to the adaptive movement strategy of species $1$. Figures~\ref{fig7a} and \ref{fig7b}
depict the mean value of $\zeta_i$ for several values of $\epsilon$ and $R$, respectively. The error bars show the standard deviation. The colours follow the scheme in Fig.~\ref{fig1}.}
  \label{fig7}
\end{figure}
%%%%%%%%%%%%%%%%%%%%%%%%%%%%%%%%%%%%%%%%%%%%%%%%%%%%%%%%%%%%%%%%%%%%%%%

%%%%%%%%%%%%%%%%%%%%%%%%%%%
\section{Selection Risk}
\label{sec6}
%%%%%%%%%%%%%%%%%%%%%%%%%%%
The adaptive movement behaviour aims to avoid regions with a high density of dominant species to minimise the chances of being eliminated in the cyclic spatial game. For this reason, we quantify the selection risk to understand how this behavioural tactic works as an organism's self-preservation strategy.
Figures~\ref{fig7a}, and~\ref{fig7b} show the average selection risks regarding the organism's responsiveness and the perception radius, respectively. The colours follow the scheme in Fig.~\ref{fig1}; the circles show the mean value while the error bars indicate the standard deviation.
Our findings show that a small responsiveness parameter ($\epsilon=1.0$) is sufficient to $\zeta_2$ to reach its minimum. The minimum $\zeta_2$ is also reached for the shortest perception radius, $R=1$. Also, the outcomes reveal that $\zeta_1$ and $\zeta_2$ reach the maximum value for $\epsilon=4.0$. But, both $\zeta_1$ and $\zeta_2$ do not vary significantly for $R\,\geq\,1$.

%%%%%%%%%%%%%%%%%%%%%%%%%%%
\section{Discussion and Conclusions}
\label{sec7}
%%%%%%%%%%%%%%%%%%%%%%%%%%%
We studied the spatial rock-paper-scissors game, where interactions among individuals are selection, reproduction, and mobility. In our stochastic simulations, the standard model is expanded by giving organisms of one out of the species the intelligence to decide their mobility based on local stimuli. Namely, individuals calculate the level of attractiveness of their neighbourhood, thus deciding to accelerate in case of being in danger or decelerate when in their comfort zone. We studied scenarios with different levels of individuals' responsiveness to the local reality - a less intense or a more abrupt individual's reaction results in a smooth or sharp speed change.
We also explored what happens if individuals have different physical abilities to perceive their local reality by implementing a perception radius. We investigated the effects of adaptive movement strategy on the individual spatial patterns, selection risk, and species densities.

Generally speaking, our results gave us robust evidence that the adaptive movement tactic does not constitute a positive strategy for a single species in a system with cyclic selection interactions. Because the species are spatially segregated in spiral waves, the alterations in the movement probability introduce a delay in the collective movement. This increases the group vulnerability to the invasion of the dominant (repellent) species and reduces the chances of individuals taking the territory occupied by the dominated (attractant) species. This happens because individuals decelerate when in the imminence of entering the areas dominated by the species they can select, thus, allowing their target species to grow. This is the reason for the shortening of the average characteristic length of single-species domains whose individuals control the mobility rate. In summary, adaptive movement strategy leads to the species' population decline; thus, fewer individuals result in more vulnerability and a consequent higher selection risk. In other words, this adapting the mobility according to the local attractiveness has been shown to be disadvantageous for the species at the individual level (selection risk grows) and at the population level (spatial density decrease). This paradoxical effect is inherent in the rock-paper-scissors models because of the cyclic dominance among the species - see for example,  Refs. \cite{dois0,dois1,dois2}.

The outcomes can be extended for spiral patterns with an arbitrary odd number $N$ of species, following a model that allows the emergence of spirals with $N$ arms.
In this generalised scenario, if organisms of species $i$, with $i=1,2,3,...,N$, evolve to move adaptively,
they benefit species $i+1$, but harm their own species. The reason is the deceleration of organisms of species $i$ when in the imminence of entering in areas dominated by species $i+1$. This reduces the chances of conquering territories for expanding their population. More, once saved for being selected, organisms of species $i+1$ multiply, reducing the number of individuals of species $i+2$.

Our study focused on the case where individuals of only one out of the species control their foraging rate according to the neighbourhood's attractiveness in the standard random walk model, widely investigated in the literature \cite{Reichenbach-N-448-1046,Szolnoki_2020}. However, this movement strategy can be investigated in scenarios of cyclic models with directional mobility. For example, it has been shown that the Safeguard directional movement brings advantage at the individual and the population level; thus, a scenario where individuals can choose the best direction to move and simultaneously adjust their mobility may optimise the results achieved in terms of territorial control. Moreover, the movement strategies may be alternated, allowing the organism to adapt to the changes of their local reality. Our discoveries may also be helpful to the understanding of more general biological systems, where learning and adaptive processes are fundamental to biodiversity stability. 
	
%%%%%%%%%%%%%%%%%%%%%%%%%%%%%%%%%%%%%%%%%%%%%%%%%%%%%%%%%%%%%%%%%%%%%%
\section*{Acknowledgments}
We thank CNPq, ECT, Fapern, and IBED for financial and technical support.

\bibliographystyle{elsarticle-num}
\bibliography{ref}

\end{document}